\documentclass[aps,prb,twocolumn,showpacs,letterpaper, superscriptaddress]{revtex4}
\usepackage{graphicx}
\begin{document}

\title{Exciton spectroscopy of hexagonal boron nitride using non-resonant x-ray Raman scattering}
\author{Yejun Feng} \email [Corresponding author. Email: ]  {yejun@aps.anl.gov}
\affiliation{Advanced Photon Source, Argonne National Laboratory, Argonne, IL 60439, USA} 
\author{J.A. Soininen}
\affiliation{Department of Physical Science, University of Helsinki, Helsinki, FIN-00014, Finland}
\author{A.L. Ankudinov}
\affiliation{Department of Physics, University of Washington, Seattle, WA 98195, USA}
\author{J.O. Cross}
\affiliation{Advanced Photon Source, Argonne National Laboratory, Argonne, IL 60439, USA}
\author{G.T. Seidler} \email [Corresponding author. Email: ]  {seidler@phys.washington.edu}
\affiliation{Department of Physics, University of Washington, Seattle, WA 98195, USA}
\author{A.T. Macrander}
\affiliation{Advanced Photon Source, Argonne National Laboratory, Argonne, IL 60439, USA}
\author{J.J. Rehr}
\affiliation{Department of Physics, University of Washington, Seattle, WA 98195, USA}
\author{E. L. Shirley}
\affiliation{Optical Technology Division, NIST, Gaithersburg, MD 20899, USA}
\date{\today}

\begin{abstract}
We report non-resonant x-ray Raman scattering (XRS) measurements from hexagonal boron nitride for transferred momentum from 2 to 9 $\mathrm{\AA}^{-1}$ along directions both in and out of the basal plane.  A symmetry-based argument, together with real-space full multiple scattering calculations of the projected density of states in the spherical harmonics basis, reveals that a strong pre-edge feature is a dominantly $Y_{10}$-type Frenkel exciton with no other \textit{s}-, \textit{p}-, or \textit{d}- components.  This conclusion is supported by a second, independent calculation of the \textbf{q}-dependent XRS cross-section based on the Bethe-Salpeter equation.  
\end{abstract}



\maketitle

The physics of low-energy photoelectrons in solids is a complex, many body problem \cite{1}.  All aspects of the electronic structure of the material must be taken into account \cite{2, 3}, including the interaction between the photoelectron and the complementary hole \cite{3,4,5,6}. This latter effect allows for similar long-lived bound states in metals (generally called Fermi-edge singularities) \cite{7} and in insulators (called excitons) \cite{8}. As a canonical example of a many body problem in condensed matter physics, core-excitons are a topic of continuing interest \cite{4,5,6}. 

The present synergy between steadily progressing \textit{ab initio} theoretical treatments \cite{2,3,9, 10} and on-going improvements in instrumentation \cite{11} in studies of non-resonant x-ray Raman scattering (XRS) shows a strong potential for rapid progress on this old problem \cite{5,6}.  XRS is the inelastic scattering of hard x-rays from bound electrons, and XRS studies have seen a recent explosion in the number and range of applications \cite{5,6,11,12,13,14,15,16,17}. In comparison with \textbf{q}-resolved electron energy loss spectroscopy (EELS) \cite{18}, XRS is more suited for bulk condensed matter systems \cite{13} due to the large penetration length afforded by the relative high-energy incident and scattered photons.  The measured \textbf{q}-dependent XRS provides direct information about multipole contributions to the dynamic structure factor $S(\textbf{q},\omega)$ \cite{5,6, 14, 15,16}.  It can be directly compared with theoretical calculations of $S(\textbf{q},\omega)$ \cite{6, 14,15,16}, and can in some cases be inverted to provide an experimental measure of the perturbed projected density of states (\textit{l}-DOS) \cite{10,19}, which can again be compared to theory.  

XRS has shown itself to be especially suitable for spectroscopy of the angular characteristics of core excitons at the near edge region.  For example, XRS studies  \cite{5,6} on LiF and icosahedral B$_{4}$C convincingly demonstrated \textit{s}-, and \textit{p}-type excitons at the F and B K-edges, respectively, which can be attributed to atoms located at a center of inversion symmetry of the unit cell with parity a good quantum number for the final states \cite{6}.  Here, we demonstrate that for some systems one can learn not only the $\Delta l$ selection rule for the exciton but also a full description of its angular characteristics in terms of spherical harmonics, $Y_{lm}$.

We present measurements of the \textbf{q}-dependent XRS of hexagonal boron nitride (h-BN), an anisotropic system with all atoms sitting at positions with reflection symmetry to the basal plane (Figure 1, inset) \cite{20}. The anisotropy of the dipole-limit core excitation spectra for $\textbf{q}\parallel c$ and $\textbf{q}\perp c $ for h-BN is well known from previous EELS\cite{21},  XANES \cite{22, 23}, and XRS \cite{24} studies at the $\textit{q}=0$ limit.  We extend on this prior work by measuring the \textbf{q}-dependent boron XRS for both   $\textbf{q}\parallel c$ and $\textbf{q}\perp c $ out to $q = 9 \mathrm{\AA}^{-1}$, a momentum transfer that is clearly beyond the dipole limit for the B 1\textit{s} initial state.  By expanding the final state wavefunction onto the spherical harmonics basis, we identify a predominantly $Y_{10}$-type exciton in the pre-edge region.  This conclusion is critically supported by two independent \textit{ab initio} calculations, one of the perturbed \textit{l}-DOS and the other a direct calculation of the \textbf{q}-dependent XRS.

\begin{figure} 
\begin{center}
\includegraphics[width=3.35in]{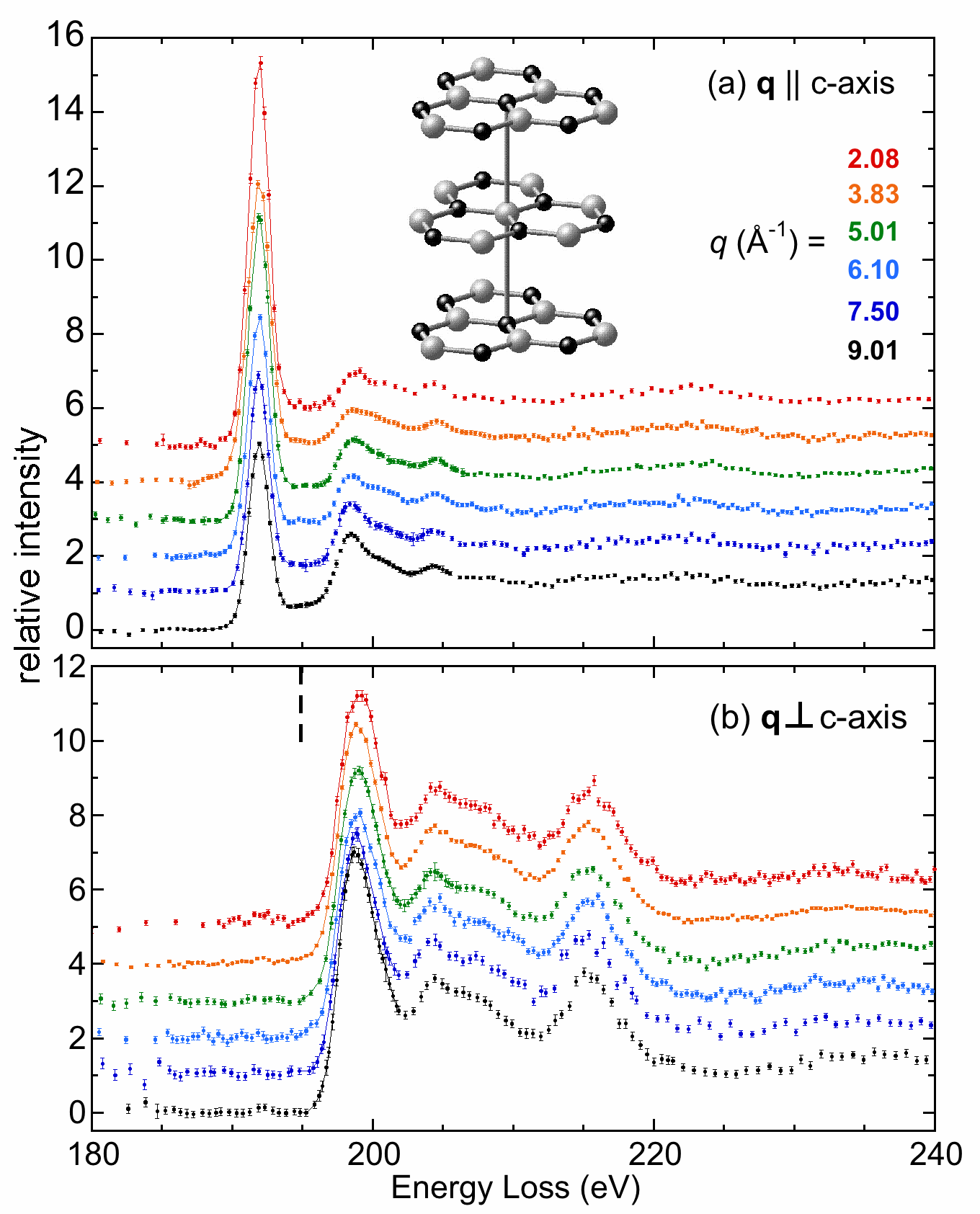}
\caption{(color online). (Inset): The crystalline structure of h-BN has 2-dimension hexagonal sheets stacked directly on top of each other, with the boron (black) and nitrogen (gray) always being nearest neighbors \cite{20}. (Main figure): The normalized XRS spectra from h-BN for $\textbf{q}\parallel c$ (top panel) and  $\textbf{q}\perp c$ (bottom panel).  The dashed line on the x-axis indicates the integration boundary for the pre-edge peak intensity shown in Fig. 4.}
\end{center}
\end{figure}

Our sample is highly-oriented compression-annealed pyrolytic h-BN (Advanced Ceramics Research) \cite{25} with dimensions $14\times5\times0.9$ mm$^3$.  The crystallite size is of the order of 1000 $\mathrm{\AA}$ with a mosaic spread about 2 to 3 degrees along the c-axis \cite{24}.  The orientation within the basal plane is assumed to be random.  All measurements were performed at the XOR/PNC-20-ID beamline of the Advanced Photon Source at Argonne National Laboratory.  An 11.4-cm-diameter spherically bent Si (5, 5, 5) crystal analyzer was mounted in the vertical scattering plane on the arm of a Huber diffractometer, forming a near-backscattering Bragg diffraction geometry with the Amptek Si-PIN photodiode detector just above the sample position. The incident x-ray energy was scanned with the x-ray analyzer energy fixed at 9890 eV.  Henceforth, we refer to all energies in terms of energy loss with respect to the elastic scattering peak.  Inelastic x-ray scattering spectra were measured at $q = 2.08, 3.83, 5.01, 6.10, 7.50$, and $9.01 \mathrm{\AA}^{-1}$ for directions parallel and perpendicular to the c-axis, with a resolution varying from 0.25 to 0.11 $\mathrm\AA^{-1}$.  Further details of our apparatus have been reported elsewhere \cite{6,26}.   All of the elastic peaks are fit well by a Gaussian function, defining the energy resolution to be between 1.32 eV and 1.66 eV FWHM with the broadening a consequence of the different effective sample size as a function of scattering angle.

\begin{figure}
\begin{center}
\includegraphics[width=3.35in]{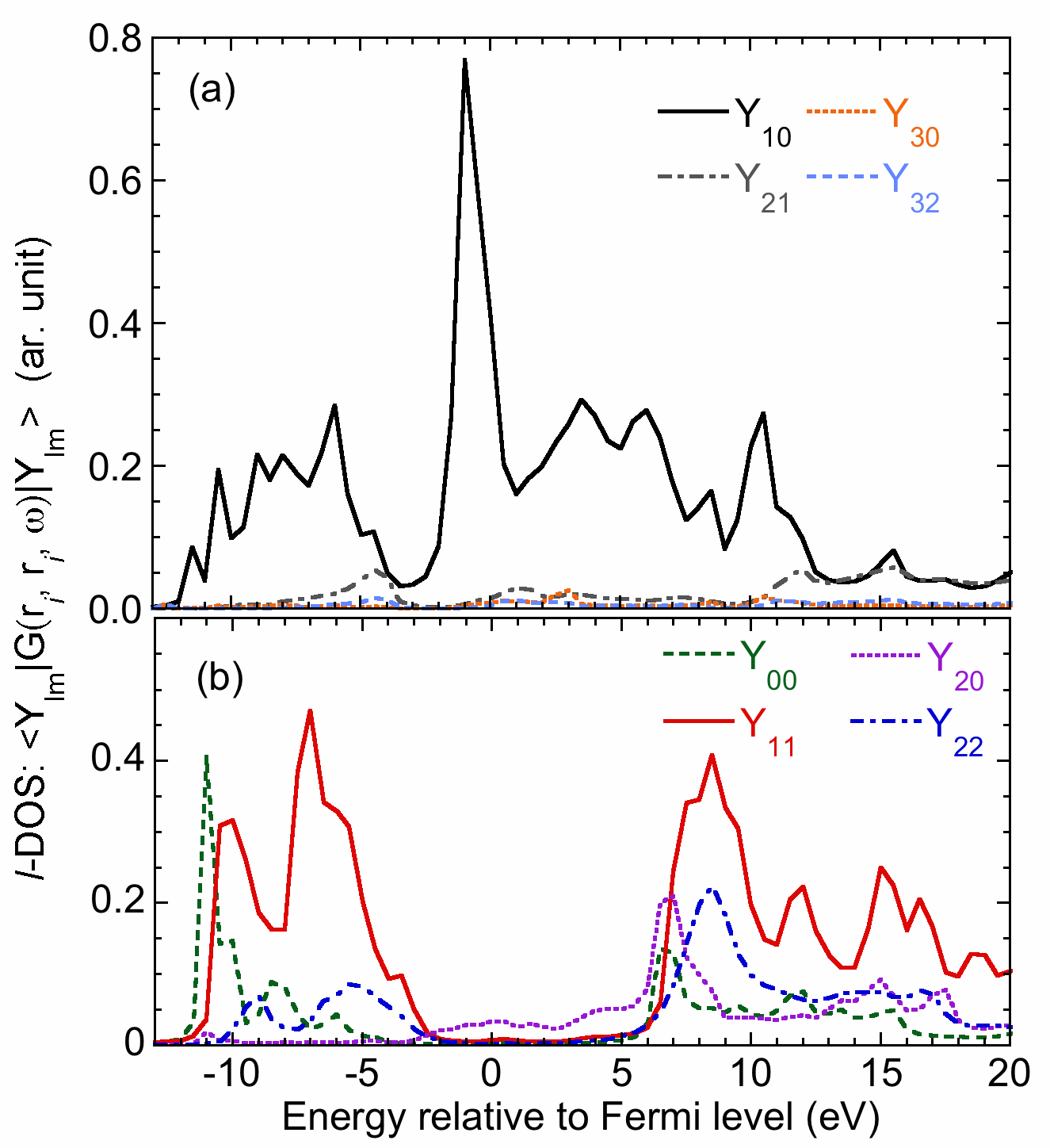}
\caption{(color online).  \textit{Ab initio} FEFF8 calculation of the \textit{l}-DOS onto different spherical harmonics with odd (top panel) and even (bottom panel) parity for an h-BN cluster of 329 atoms.  See the text for details.}
\end{center}
\end{figure}

The separation of the XRS signal from the larger Compton background is a key experimental difficulty in all XRS measurements \cite{16,27}.  The valence Compton backgrounds are approximated by fitting the pre-edge data to standard Compton forms and using the low-$q$ limit XRS form as a reference \cite{26}.  The data were numerically broadened, as necessary, to the maximum resolution of 1.66 eV FWHM for visual comparison.  

The resulting boron XRS spectra are presented in Figure 1.  The pre-edge peak position is $191.9\pm0.1$ eV for the c-axis spectra, and the main edge peak position is $198.8 \pm 0.1$ eV along both directions, consistent with prior work using other experimental methods \cite{21,22,23,24}. All of the spectra were normalized according to the integrated intensity in the 210 eV to 250 eV range, mimicking the $f$-sum rule, and naturally introducing a $q^2$ factor\cite{28}. The resulting XRS spectra are shown in Fig. 1 for both $\textbf{q}\parallel c$ (top panel) and $\textbf{q}\perp c$ (bottom panel).

\begin{figure}
\begin{center}
\includegraphics[width=3.35in]{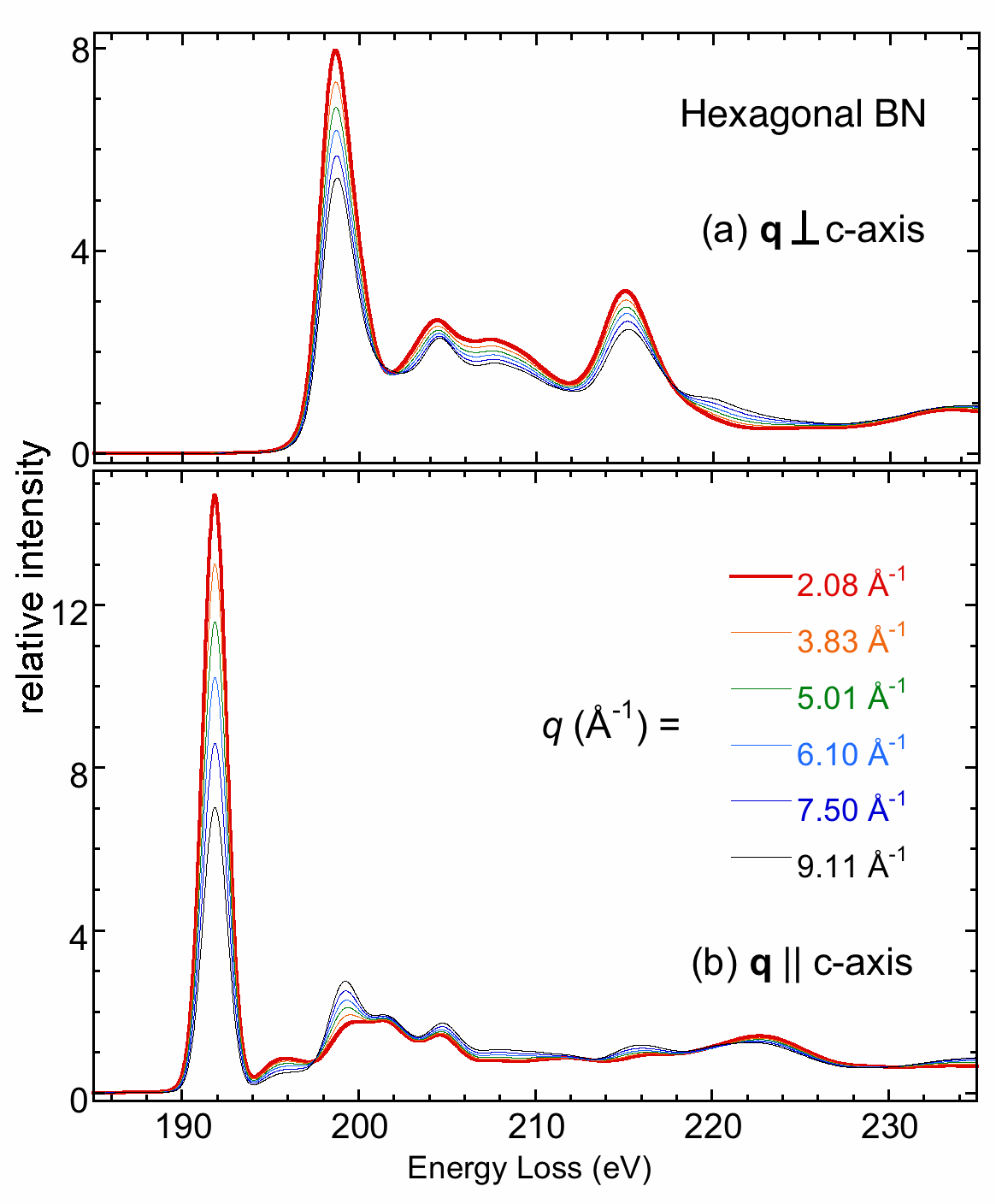}
\caption{(color online). First-principles calculation of  $S(\textbf{q}, \omega)$ over the experimentally measured \textit{q} range along both (a) $\textbf{q}\perp c$ and (b) $\textbf{q}\parallel c$ directions.  See the text for details.}
\end{center}
\end{figure}

Previous experimental studies have suggested that the physical origin of the pre-edge peak in h-BN is a core exciton\cite{22,29}. Theoretical calculations have confirmed this notion by demonstrating the importance of the interaction between the core-hole and the photoelectron \cite{30}.  In the XRS spectra for $\textbf{q}\parallel c$ (Fig. 1, top panel) this exciton state exhibits a prominent monotonically decreasing peak intensity with increasing $q$.  Another less obvious but equally significant feature is the total absence of the exciton peak in spectra measured within the basal plane for any $q$ values.  Although the shape and intensity of the main edge peaks along both directions also display subtle variations, we concentrate on the behavior of the exciton.

We now present a general formalism for \textbf{q}-dependent XRS in order to understand the interplay between the final-state characteristics measured by core-exciton spectroscopy and the local symmetry at the sites of interest.  The dynamic structure factor $S(\textbf{q}, \omega)$ is expressed in a single-electron picture as 
\begin{equation}
S(\mathbf{q}, \omega)=\sum_f\vert \langle f \vert e^{i\mathbf{q}\cdot \mathbf{r}}\vert 0\rangle\vert^2 \delta(\Delta E-\hbar\omega).
\end{equation}
Following the short-range order theory \cite{32} due to the localized core-hole effect and the general guideline of Doniach \textit{et al}. \cite{12},  the angular parts of both the electron wave functions and transition matrix element are expanded in the spherical harmonics basis as
\begin{equation} 
\psi_f(\mathbf{r})=\sum_{l'm'}R^{f}_{l'm'}(r)Y_{l'm'}(\theta, \varphi)
\end{equation} 
and 
\begin{equation}
e^{i\mathbf{q}\cdot \mathbf{r}}= 4\pi \sum_{l,m} i^{l} j_{l}(qr) Y_{lm}(\theta_{q},\varphi_{q})Y^{*}_{lm}(\theta, \varphi),
\end{equation} 
with $R^f_{l'm'}(r)=\langle Y_{l'm'}\vert f\rangle$.  The initial core state is assumed to be $s$-type, and the final state is usually a mixture of partial waves because of the ligand-field influence in a solid system.  Substitution of these expansions into Eq. 1 separates the angular variables of the transferred momentum $\textbf{q}$ from its magnitude in $S(\mathbf{q}, \omega)$ as
\begin{equation}
S(\mathbf{q}, \omega)=\sum_f\arrowvert \sum_{lm} s_{flm}(q)Y_{lm}(\theta_q, \varphi_q) \arrowvert^2 \delta(\Delta E-\hbar\omega),
\end{equation}
where 
\begin{equation}
s_{flm}(q)=i^l \sqrt{4\pi} \int_{0}^{\infty} j_{l}(qr)R^{f}_{lm}(r)R_{0}(r)\cdot r^2 dr
\end{equation}
is the magnitude of the $Y_{lm}$-projected momentum space convolution between the initial state $\vert 0 \rangle$ and each final state $\vert f \rangle$ \cite{26}. This approach is very similar to the formalism of Soininen \textit{et al.}\cite{10}, but without the directional averaging (\textit{i.e.}, summation over \textit{m}) employed in that work. 

\begin{figure}
\begin{center}
\includegraphics[width=3.35in]{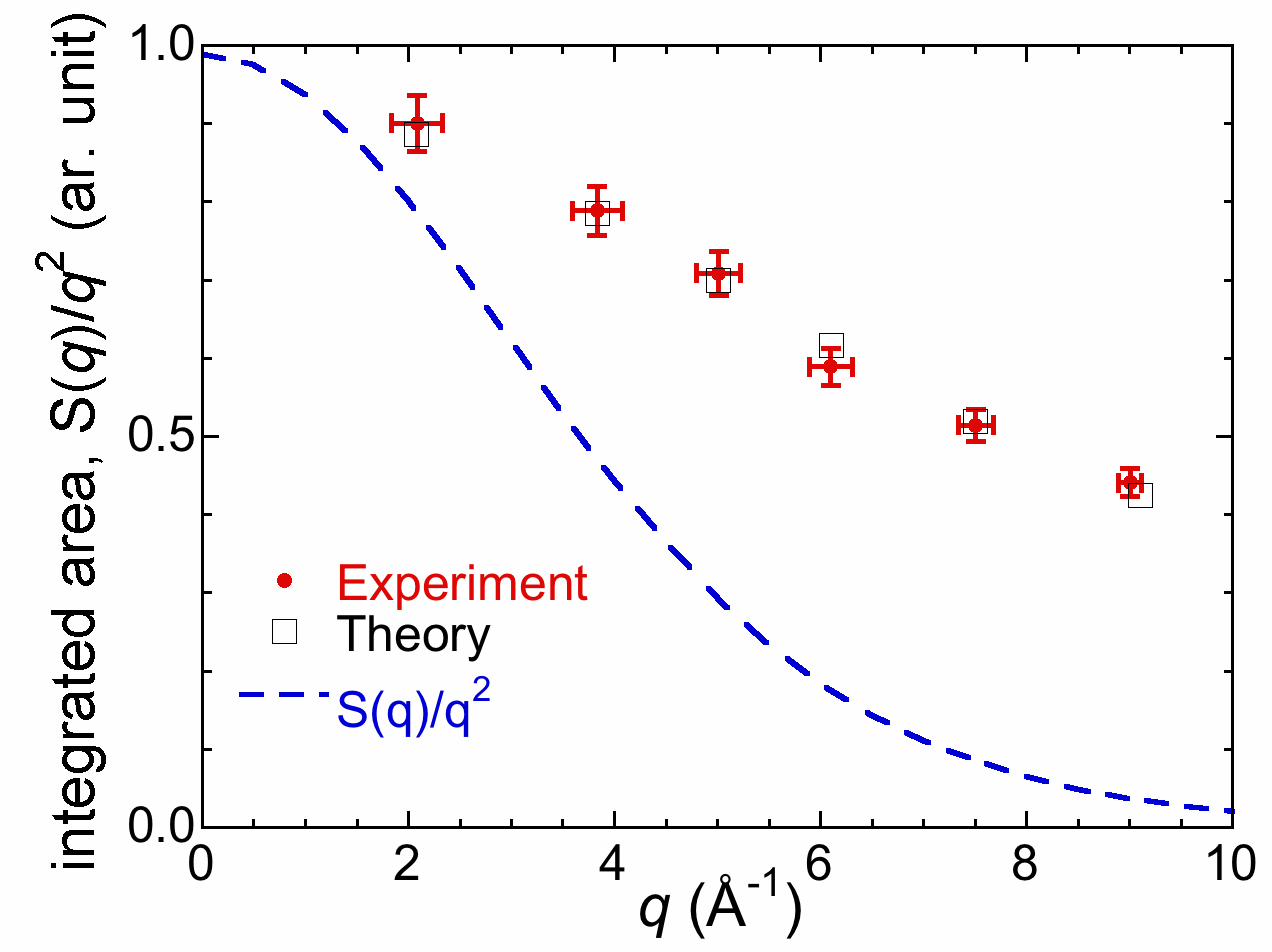}
\caption{(color online). (solid dots with error bars, red): Integrated exciton intensities from the experimental XRS data \label{Fig1} for energies below 194.5 eV.  (open squares, black): Theoretically calculated exciton intensity (Fig. 3) integrated over the same energy range and plotted with an overall adjustable scaling factor to the experimental data.  (dashed curve, blue): A simple calculation of $S(q)/q^2$ using boron 1\textit{s} core-state and a crude hydrogen-like $2p_z$ final state with a $Z_{boron}+1$ approximation for the core-hole\cite{6,26}. }
\end{center}
\end{figure}

For the first three orders of spherical harmonics expansion (\textit{l} up to 2), partial waves of $Y_{00}$, $Y_{1,\pm 1}$, $Y_{20}$, and $Y_{2,\pm 2}$ could all have non-vanishing contributions to $S(\textbf{q}, \omega)$ for $\textbf{q}\perp c$.  Hence, the non-existence of the exciton peak for $S(\textbf{q}, \omega)$ measured along $\textbf{q}\perp c$ for any magnitude of \textit{q} indicates the absence of these partial waves.  Also the $Y_{2,\pm1}$ do not have contributions to the XRS spectra along either direction. This leaves $Y_{10}$ the only spherical harmonic component for $l<3$ to contribute to the $\textbf{q}\parallel c$ spectra. 

This identification of a $Y_{10}$-type exciton is consistent with the fact that boron atoms sit only at sites with reflection symmetry about the \textit{a-b} basal plane, which demands that each final state be an eigenstate of the corresponding reflection parity \cite{22}. For a predominantly $Y_{10}$-type exciton, other possible higher-order $Y_{lm}$ components must possess the same reflection parity.  This symmetry argument puts stringent restrictions on other possible spherical-harmonic components to the final state wave function. Note that $Y_{00}$, $Y_{1,\pm1}$, $Y_{20}$, and $Y_{2,\pm2}$ all have even parity about the \textit{a-b} plane, while only $Y_{10}$ and $Y_{2,\pm 1}$ have odd parity.  Given the additional reflection symmetry about the \textit{a-c} plane for boron sites (Fig. 1, inset), $Y_{2,\pm 1}$ could be also excluded from the final state wave function, because of the different \textit{a-c} plane reflection parity in comparison with $Y_{10}$.  Thus for $l<3$, the exciton state has a purely $Y_{10}$ type of wave function. This is our central result.

Compared to previous anisotropy studies by XANES and EELS that only excluded $Y_{1,\pm 1}$ contributions because of the dipole selection rule\cite{21}, the present \textbf{q}-dependent XRS study provides a more complete picture of the exciton state characteristics as of dominantly $p_z$-type ($Y_{10}$), with no $s$-, $p_{x,y}$-, or $d$-components.  The above analysis, in which the  \textbf{q}-dependent XRS spectra may be qualitatively interpreted in the context of the constraints imposed by the crystal symmetry, should be generically applicable to the analysis of the angular characteristics of excitons in other high-symmetry crystals.

Seeking independent verification and further insight into the exciton's $\textbf{q}$-dependence, we performed two independent first principles theoretical calculations.  First, we modified the FEFF8 \cite{33} code to calculate the density of final states projected onto the individual spherical harmonics as $\langle Y_{lm}\vert G(\mathbf{r}, \mathbf{r}, \omega)\vert Y_{lm}\rangle$, with the core-hole effect included (Figure 2). The calculation confirms our parity conservation argument by showing no contribution from even reflection parity $Y_{lm}$'s and also indicates the \textit{l}-DOS for other odd $Y_{lm}$'s with \textit{l} up to 3 are negligibly small.  This includes $Y_{30}$, which is the only other possible contributor to the $\mathbf{q}\parallel c$ spectra for \textit{l} up to 3.  Note that the single $Y_{10}$ characteristic of the wave-function indicates the exciton state is localized at the scattering atom. Hence it could not be a superposition of 2\textit{p} states from both the excited boron atom and its neighboring nitrogen atoms, as typically in either a band structure hybridization or the molecular orbital theory. Such a superposition of wave functions from different atomic sites would not generate a single spherical harmonic wave function. This is a significant refinement of the prior understanding of this exciton peak as a $1s-\pi^*$ transition \cite{21,24, 34}.

Second, we calculated the XRS spectra from the boron K-edge using a computational Bethe-Salpeter equation method for modeling core-excited states\cite{31}, while including calculation of the Hedin-Lundqvist's GW self-energy\cite{1} for both the energy dependent quasi-particle shifts and the quasi-particle lifetime broadening.  Convergence was checked with respect to the numerical cutoffs\cite{35}. The simulated spectra along both axes are plotted in Figure 3, using the same normalization and resolution as the experimental data.  The simulations also verify that the general spectral profile far away from the near edge has little variation in the probed $q$ range.

We plot in Figure 4 the integrated exciton intensity for energies below 194.5 eV as a function of $q$ for $\mathbf{q}\parallel c$ for both experimental data (solid dots with error bar) and the K-edge calculation (open squares).  The estimated relative standard uncertainty of the integrated intensity is 4\% from a combination of statistical and estimated systematic errors applying to each independent measurement.  Due to the weak \textit{q}-dependence of the atomic background \cite{10}, our method of normalizing the spectra may also introduce another 20\% systematic error into the overall, smooth \textit{q}-dependence of the integrated area under the pre-edge feature.  Hence the agreement between theory and experiment is indeed good, but certainly not so exacting as is shown in Fig. 4. As the theory has no lattice motion (phonon) included, the Frank-Condon effect from electron-phonon coupling \cite{4,36} is not critical and the momentum is largely conserved during scattering\cite{37}. 

For XRS spectra for $\mathbf{q}\perp c$, the experimental results and first principles calculations have a good qualitative agreement on both the general spectral structure and peak positions (Fig. 3).  However, there is a disagreement about the $q$-dependence of the main edge peak at 199 eV.  The experimental data shows about 15\% increase in this peak's intensity when progressing from 2 to $9 \mathrm{\AA}^{-1}$, while the theory predicted a decreasing trend. The discrepancy is too large to be accounted for by the randomly oriented h-BN crystallites within the basal plane, and may be partially mediated by a weak \textit{q}-dependence in the atomic background \cite{10}. Future studies are needed to clarify it.  However, this discrepancy does not impact our main results.

We also performed a crude model calculation of this exciton's $q$-dependence using a hydrogen-like $2p_z$ final state with a screened (static dielectric constant $\epsilon=7$) $Z+1$ approximation of the core-hole potential from Eq. 1 (dashed line in Fig. 4)\cite{6,26}, resulting in a $q$-dependence much narrower than the experimental data.  The failure of this calculation suggests a deeply-bound Frenkel-type exciton\cite{8}, which would provide a broader final momentum distribution.

In conclusion, we have used non-resonant x-ray Raman scattering (XRS) to perform exciton spectroscopy on h-BN, and find a $Y_{10}$-type exciton at the pre-edge position of the boron K-shell spectra for $\mathbf{q}\parallel c$.  This conclusion is supported by independent first principles calculations of the XRS and of the projected density of states.  The existence of a pure-type exciton is strongly related to the presence of reflection symmetry in the system.  This work illustrates the role of symmetry-based arguments in the interpretation of single-crystal \textbf{q}-dependent XRS studies and more broadly illuminates the interplay between local atomic symmetry and final-state effects in core-shell spectroscopies.

We acknowledge the stimulating discussions with E. A. Stern, E. A. Miller, and B. D. Chapman.  J.A.S. acknowledges financial support from the Academy of Finland under Contract No 110571/1112642. This research was also supported by the L. X. Bosack and B. M. Kruger Charitable Foundation; the Mellam Family Foundation; US DOE grants DE-FGE03-97ER45628, and DE-FG03-97ER45623 facilitated by the CMSN; and the Natural Sciences and Engineering Research Council of Canada.  Use of the Advanced Photon Source is supported by the U.S. DOE-BES, under Contract No. DE-AC02-06CH11357.


\begin{thebibliography}{50}
\bibitem {35} Spectra were calculated using 60 conduction band states with an 81 Ry plane-wave cutoff and 4096 k-points.
\end{thebibliography}
\end{document}